# Application of FPGA Acceleration in ADC Performance Calibration

Guangyuan Yuan, Zhe Cao, Shuwen Wang, Shubin Liu, Qi An

*Abstract*—In recent years, high speed and high resolution analog-to-digital converter (ADC) is widely employed in many physical experiments, especially in high precision time and charge measurement. The rapid increasing amount of digitized data demands faster computing. FPGA acceleration has an attracting prospect in data process for its stream process and parallel process feature. In this paper, an ADC performance calibration application based on FPGA acceleration is described. FPGA reads the ADC digitized data stream from PC memory, processes and then writes processed result back to the PC memory. PCIE bus is applied to increase the data transfer speed, and floating point algorithm is applied to improve the accuracy. The test result shows that FPGA acceleration can reduce the processing time of the ADC performance calibration compared with traditional method of C-based CPU processing. This frame of PCIE-based FPGA acceleration method can be applied in analysis and simulation in the future physical experiment for large ADC array, such as CCD camera and waveform digitization readout electronics calibration.

## I. Introduction

ADC (analog-to-digital converter) plays an important role in physics experiments in analog signal acquisition [1,2]. The performance of ADC is very important for it influences the data accuracy and experiment result. Thus the calibration of ADC performance is very important. The most concerned dynamic parameters of an ADC chip are SNR (Signal-to-Noise Ratio), SINAD (Signal to Noise and Distortion Ratio), ENOB (Effective Number of Bits), THD (Total Harmonic Distortion) and SFDR (Spurious Free Dynamic Range). FPGA (Field-Programmable Gate Array), for its parallel processing and stream processing structure, has an attracting feature in acceleration of data process [3,4]. The data stream can be transferred into FPGA, processed, and write back to PC memory simultaneously. The data processing speed can reach up to 2 GB/s in design in FPGA. PCIE-bus based DMA (Direct Memory Access) was applied in data transmission between FPGA and PC, and floating point was applied to improve the accuracy.

## II. Principle and method

The dynamic parameters are all calculated on the base of spectrum analysis, mainly consisting of FFT (Fast Fourier Transform), which is very time consuming [5]. As FFT is calculated parallel, meeting the feature of FPGA computing, so it is suitable to be separated and processed by FPGA. After spectrum analysis, FPGA write results back to the PC memory and the program continues to calculate the parameters of ADC.

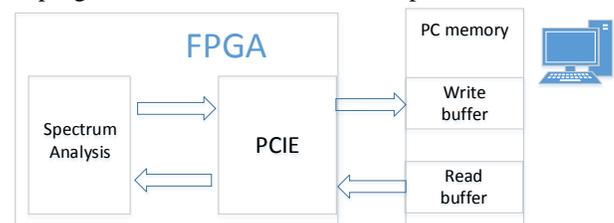

Figure 1. Structure of data stream

The structure of data process stream is shown in Fig.1. FPGA reads data from PC memory, implements spectrum analysis and then writes results back to PC memory.

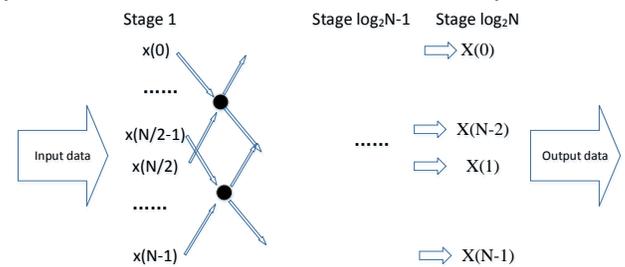

Figure 2 pipelined stream process.

For an N points FFT, the amount of complex multiplication is $N \times \log_2 N$. In the FPGA, N lines can run simultaneously, $\log_2 N$ level was need for an N-point group. For a signal 1024-point FFT, 4258 clocks are needed. However, as the FPGA can be configured to work in pipeline, multi-level operation can be carried out simultaneously. That is, multiple group data can be processed in the same time. As shown in Fig. 2, while data group 1 is on Stage $\log_2 N$, data group 2 is being processed in stage $\log_2 N-1$, and data group N is streamed in on stage 1. The feature of parallel process and stream process result in acceleration in process speed and reduction in time.

Manuscript received June 9, 2018. This work was supported by the Key Program of the National Natural Science Foundation of China (Grant No. 11635007), the Young Fund Projects of the National Natural Science Foundation of China (Grant No. 11505182), and the State Key Laboratory of Particle Detection and Electronics (No. SKLPDE-ZZ-201814).

Authors are with the Department of Modern Physics, University of Science and Technology of China, Hefei, Anhui 230026, China. Corresponding author: caozhe@ustc.edu.cn.

## III. VERIFICATION TEST

The performance test of the design was carried on in Virtex-7 XC7VX690T (Xilinx FPGA) and xeonE5-2609 v4 (Intel CPU). Programs are designed with and without FPGA separately to compare the results.

A group data, consist of 1024 64-bit points, was generated to test the performance of the programs. The data was generated by formula:

$X(t)=0.7*cos(2*pi*50*1000000*t)+sin(2*pi*12*1000000*t)+0.1*randn()$.

In program applying FPGA acceleration, one group data was carried into FPGA at one time. After processing, the FPGA wrote data back to PC memory and the program continued to calculate the parameters.

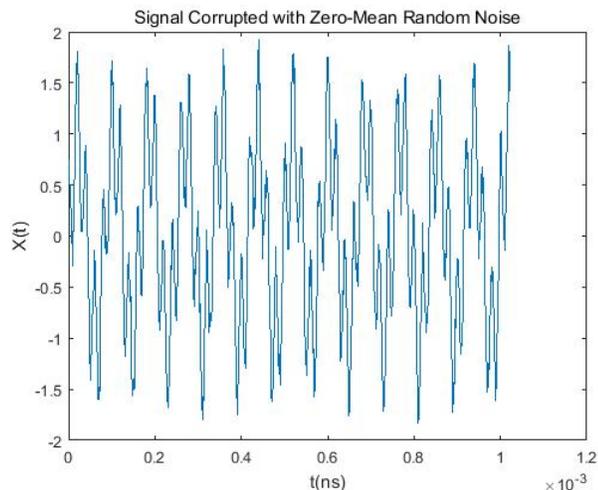

Figure 2. Time domain signal with random Noise.

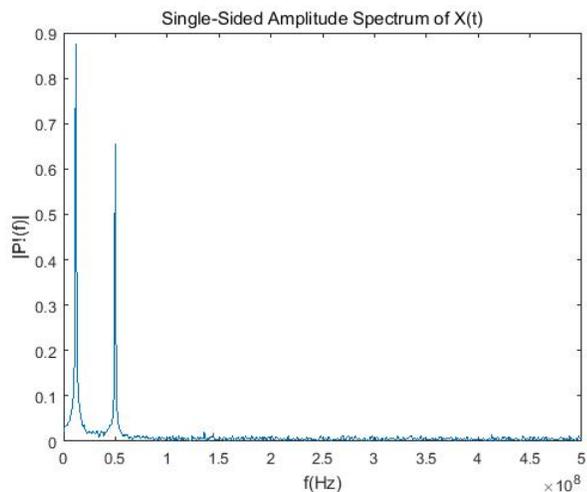

Figure 3. Frequency domain signal after processing.

Figure 2 shows the data generated by PC, consisting of a 50 MHz signal, a 12 MHz signal and random noise. Figure 3 is the result shown in frequency domain after process. The results applying FPGA and pure CPU processed are the same, meeting design expectations.

CPU runs in 1.7 GHz while the system of FPGA runs in 250 MHz. 10^9 times are repeated to figure out the average process time of the programs. The average time processed by program written in C++ is 2.19 ms, while that of program applying FPGA is 18 us, consistent with the expected result. As a group data consist of 1024 64-bit points, the corresponding process speed is 440MB/s. The speed is fastened by more than 100 times.

## IV. DISCUSSION

If continuous data are provided to spectrum analysis rather than one group data once, the average processing time can reach up to the limit of FFT bus speed, which is 2 GB/s when the system clock is designed to 250 MHz. Besides, more FFT module can be implemented in FPGA to further accelerate the speed to the limit of PCIE bus.

In addition, other particular data process can also be accelerated by FPGA. The calculation of ENOB, SNR, SINAD, THD, SFDR and SFDR can also be done in FPGA.

## V. CONCLUSION

A FPGA-based acceleration structure was introduced, spectrum analysis for ADC calibration was performed and system performance was tested in our work. PCIE bus and floating point algorithm were applied to improve the performance. The test results shows the speed can be fastened 100 times by FPGA acceleration, while the accuracy was the same. The result shows FPGA acceleration has an attracting feature in physical experiment data processing. Besides, the implement of spectrum analysis in FPGA can be applied in on-board calibration for large ADC array, such as CCD camera calibration.

### REFERENCES

[1] R.H. Walden, "Analog-to-digital converter survey and analysis", IEEE Journal on Selected Areas in Communications, vol. 17, pp. 539-550, Apr. 1999.
[2] R.H. Walden, "Performance trends for analog to digital converters", IEEE Communications Magazine, vol. 37, pp. 96-101, Feb. 1999.
[3] A, Gothandaraman, et al. "FPGA acceleration of a quantum Monte Carlo application", Vol. 34, pp.278-291, May. 2008.
[4] J.P. Walters, et al. "MPI-HMMER-Boost: Distributed FPGA Acceleration", The Journal of VLSI Signal Processing Systems for Signal, Image, and Video Technology.Vol. 48, pp. 223-238, Sep. 2007.
[5] W.T. Cochran, et al. "What is the fast Fourier transform?", Proceedings of the IEEE, Vol. 55, pp. 1664-1674, Oct. 1966.